# La-doped $CH_3NH_3BaI_3$ : A Promising Transparent Conductor

**Jiban Kangsabanik[#], Vikram[#], Aftab Alam[*]**

*Department of Physics, Indian Institute of Technology, Bombay, Powai, Mumbai 400076, Maharashtra, INDIA.*

[*]Corresponding Author : (+91) 22 25765564, aftab@iitb.ac.in
[#]These authors contributed equally to this work

## ABSTRACT

Hybrid perovskites ($CH_3NH_3PbI_3$) is one of the most promising novel materials for solar harvesting. Toxicity of lead (Pb), however, has always remained a concern. We investigated the electronic structure of complete replacement of Pb by alkaline earths (Ca, Sr, Ba) and found it to be wide band gapped ($E_g$) semiconductors (band gap ~ 3.7 to 4.0 eV), and hence not suitable as absorber material. This opens up a new avenue to explore these materials as transparent conductor (TC). We doped $CH_3NH_3BaI_3$ (largest $E_g$) with La, which shifts its Fermi level ($E_F$) at conduction band bottom and induces states at $E_F$ for conduction. This is precisely what is required for a transparent conductor. Optical and transport properties simulated from linear response (within Density Functional Theory (DFT)) calculations suggested it to be a very good TC material with a high figure of merit ($\sigma/\alpha$), where $\sigma$ is the electrical conductivity and $\alpha$ is the optical absorption coefficient. This claim is also supported by our calculated results on density of states at $E_F$, effective mass, carrier concentration etc. at various La-doping. We propose $CH_3NH_3(Ba_{1-x}La_x)I_3$ (x≤12.5%) to be a good TC material to be used in a all perovskite solar cell.

# Introduction

A transparent conductor (TC) is a material with high electrical conductivity as well as high optical transparency in the visible region. They have a wide range of applications in photovoltaic, display technologies, smart windows, solid-state light industries etc. There are various suitable TC's available for applications in different areas e.g. Indium tin oxide, doped ZnO, Transparent polymers, graphene, Carbon nanotube etc to name a few. To design a good TC the main requirement is to minimize the photon absorption and reflection keeping the carrier concentration high.[1]

The industrially used procedure to make a TC is to degenerately dope a wide band gap (>3.25 eV) semiconductor.[2] Organometallic lead halide perovskites ($CH_3NH_3PbI_3$) are very well known material which has application as solar absorber material with the efficiency being increased from 3.8% to ~20% in only six years.[3] But the toxicity of lead (Pb) puts a limitation on its industrial use. There are various on-going research to replace Pb with suitable elements without affecting its efficiency substantially. Alkaline earth metals (Ca, Sr, Ba) having +2 oxidation state, co-ordination number 6 and similar crystal radius are investigated as a possible replacement of Pb. But due to their lower electronegativity, they turned out to be wide band gap (>3.5eV) semiconductors, restricting their use as a solar absorber material in the visible spectrum.[4]

In this letter, we propose $CH_3NH_3BaI_3$ ($MABaI_3$, largest band gap ~ 4eV) as a possible host material, which can be doped, to be used as transparent conductors. We first show the electronic structure and optical properties of the parent material itself. Then we briefly discuss the effect of doping by various elements (namely Al, Ga and La) on its electronic properties. Detailed simulation of electronic, optical and transport properties indicated that La-doped $MABaI_3$ is a promising transparent conductor. This paper is organized as follows: computational details, Result & discussions (electronic structure of doped $MABaI_3$ with different dopants, Optical properties of pure and La doped $MABaI_3$, electrical conductivity with different concentration of La), Conclusions.

## Computational Details

*Ab-initio* calculations were performed using Density Functional Theory (DFT) with plane wave basis set as implemented in Vienna Ab initio Simulation Package (VASP).[5-7] We used Projector Augmented Wave (PAW) pseudo-potentials with Perdew–Burke–Ernzerhof (PBE) exchange correlation functional.[8,9]

We used, C 2s2/2p2, H 1s1, N 2s2/2p3, Ba 5s2/5p6/6s2, I 5s2/5p5, Al 3s2/3p1, Ga 3d10/4s2/4p1 and La 5p6/6s2/5d1 electrons as valence electrons. A plane wave energy cut off of 500 eV with a gamma-centered k-mesh was used to sample the Brillouin zone. Forces (energies) were converged to values less than 0.01 eV/Å ($10^{-6}$) eV). The Brillouin zone sampling was done by using 3x3x2 K-point mesh for relaxation and 12x12x8 for static calculations.

For La doping at Ba sites, 2x2x2 (384 atoms), 2x2x1 (192 atoms) and 2x1x1 (96 atoms) super cells were used and one Ba atom was replaced with La to achieve 3.125%, 6.25% and 12.5 % La-doping respectively. For Al & Ga doping only 6.25% doping is considered.

Optical properties are calculated within the independent particle approximation as implemented in VASP. For the calculation of electrical conductivity we solved the Boltzmann transport equation as implemented in BoltzTraP code.[10]

The following equation is used to calculate the carrier concentration (n),

$$n = \int D(E)f(E)dE = \int D(E)\frac{1}{e^{(\mu-E)/k_BT}+1}dE$$

where D(E) is the density of states, f(E) being the Fermi-Dirac distribution function for electrons of energy E. At a fixed temperature (T), the integral is evaluated over entire energy range for a given chemical potential (μ) to give the corresponding carrier concentration value. Here the chemical potential is taken at the Fermi level.

The effective mass at Gamma point is calculated using finite difference method as implemented in Effective Mass Calculator.[11]

## Results & Discussions

Our initial motivation was to check whether the replacement of toxic element Pb in MAPbI$_3$ by alkaline earth element (Ba, Ca, Sr) can still give us a useable solar material. Our ab-initio simulation revealed that, all the three alkaline earth based compounds has large band gap ($E_g$=4.0, 3.83 and 3.99 eV for $MABaI_3$, $MACaI_3$, and $MASrI_3$ respectively). This makes them unsuitable as solar harvesting material. However, it opens up the possibility to explore these materials as a transparent conductor (degenerate semiconductor). With the largest band gap, we chose $MABaI_3$ to explore further.

To design a good TC material, a reliable figure of merit is ($\sigma/\alpha$), where $\sigma$ is the electrical conductivity and $\alpha$ is the optical absorption coefficient.[3] First principles linear response calculation shows that La-doped $MABaI_3$ has low absorption co-efficient in visible spectra range. The empirical formula $\sigma=(ne^2\tau)/m_e^*$ directs us to choose a material with lower carrier effective mass (dispersive band structure) and a suitable dopant which can give appreciable carrier concentration. We first calculated the carrier effective mass of pure $MABaI_3$ at $\Gamma$ point in reciprocal space. It has lower effective mass at the conduction band edge than at the valence band edge. This suggests that the parent material can be a good n-type transparent conductor. So we decided to dope it with donor atom and hence chose to replace Ba atom with an atom with charge state 3+. We tried three different dopants starting from lighter Al, Ga and then La which is closest to Ba in periodic table. To see the effects of these doping on the electronic structure of $MABaI_3$, we first calculated the total and projected density of states.

## (a) Electronic structure

Figure 1 shows the total and Z-projected density of states for pure and $MA(Ba_{93.75}Z_{6.25})I_3$ (Z=Al, Ga, La) compounds. The valence band and conduction band for the pure $MABaI_3$ are mainly contributed by I-p orbitals and Ba-d orbitals respectively. For Al and Ga doping, new states are induced at Fermi level ($E_F$) in between the valance and conduction band, thereby reducing the effective band gap to 1.57 eV and 0.71 eV respectively. These occupied states at $E_F$ in both these cases arise mainly from the s-orbitals of the respective dopant elements and are responsible for the increase in the conductivity value. However for La doping, states arises at conduction band edge, keeping the effective band gap ~3.8 eV which is comparable to the parent material (~4.0 eV). Here La-d orbitals are the main contributors. As such, we decided to choose La as possible dopant and calculated the transport and optical properties of the La doped materials.

We chose three different concentrations of La doping (3.125%, 6.25% and 12.5%) and calculated the respective degenerate band gap ($E_g$), carrier concentrations (n) and carrier effective mass ($m_e^*$) at Γ point. These are shown in Table I.

**Table I :** Degenerate band gap ($E_g$), carrier concentration (n) and effective mass ($m_e^*$) for parent and La-doped $MABaI_3$.

| Doping % | 0% | 3.125% | 6.25% | 12.5% |
|---|---|---|---|---|
| $E_g$ (eV) | 4.00 | 3.75 | 3.84 | 3.81 |
| n ($cm^{-3}$) | 0 | $2.51 \times 10^{20}$ | $4.15 \times 10^{20}$ | $9.02 \times 10^{20}$ |
| $m_e^*$ (in units of $m_e$) | 0.425 | 15.75 | 3.83 | 3.67 |

Notably, with increase in La doping the degenerate band gap almost remains the same whereas carrier concentration increases almost linearly due to the increase in number of states at $E_F$. Carrier effective mass ($m_e^*$), however, has a non-monotonous behavior. Interestingly, the ratio ($n/m_e^*$) keeps increasing with increasing the La doping which is a promising trend for the electrical conductivity ($\sigma \propto (n/m_e^*)$).

Figure 2 shows the total and La-projected projected density of states $MABa_{1-x}La_xI_3$ at various La concentrations. It is clear from the figure that the Fermi level shifts from the valance band maxima to conduction band minima with increasing La doping. With increasing La doping, the La states go deeper into the conduction bands and the magnitude of occupied states at $E_F$ increases causing an enhancement in the conductivity.

## (b) Transport and optical properties

Within a non-varying relaxation time approximation, we calculated the electrical conductivity ($\sigma/\tau$) by solving the Boltzmann transport equation at 12.5% La doping. For other concentrations (3.125% and 6.25%), we used the conventional rigid band approximation scheme to calculate the same. Figure 3 shows the electrical conductivity vs. La-doping calculated directly using the Boltzmann transport equation. ($\sigma/\tau$) calculated this way are more reliable than the simple empirical formula $\sigma = (ne^2\tau)/m_e^*$.

We have also calculated ($\sigma/\tau$) using the empirical formula, which gave an order of magnitude higher values. This is expected because in this calculation, carrier effective mass at Γ point alone is taken into account. These values are listed in Table II, and should only be used to analyze a qualitative trend of electrical conductivity vs. La doping (x).

**Table II :** Electrical conductivity (σ/τ) vs. La-doping (x) in $MA(Ba_{1-x}La_x)I_3$, using empirical formula $\sigma=(ne^2\tau)/m_e^*$.

| La doping x (%) | 3.125% | 6.25% | 12.5% |
|---|---|---|---|
| σ/τ (ohm$^{-1}$ m$^{-1}$ s$^{-1}$) | 11x10$^{17}$ | 73x10$^{17}$ | 141x10$^{17}$ |

Finally, we compared the optical properties of Pure and 12.5% La doped MABaI$_3$. As mentioned before, absorption coefficient is another quantity, which control the figure of merit for a TC. Figure 4 shows the absorption coefficient for pure and 12.5% La doped MABaI$_3$. Absorption coefficient in the present case turns out to be comparable to those of undoped ZnO,[12] which is another well-studied material for TC applications. Notably, the absorption coefficient for the doped compound is slightly higher than the parent material, although the magnitude still remains small. Inset of Fig. 4 shows a closer look of the various peaks in the absorption co-efficient. Most importantly, the peak at/near ~310 nm corresponds to the approximate value of band gap of pure and 12.5% La doped MABaI$_3$.

## Conclusion

Lead being toxic, we investigated the replacement of Pb by alkaline earth elements (Ba, Ca, Sr) in CH$_3$NH$_3$PbI$_3$. All these compounds, CH$_3$NH$_3$YI$_3$ (Y=Ba, Ca, Sr), turn out to be large band gap (E$_g$>3.8 eV) semiconductors and hence not suitable for solar harvesting. CH$_3$NH$_3$BaI$_3$ has the largest band gap and hence chosen to dope with Al, Ga and La for possible transparent conducting properties. Al and Ga introduces donor levels in the middle of the band gap region whereas doping with La shifts the Fermi level near conduction band and introduces states at E$_F$ which account for the monotonic increase in the electrical conduction. La-doped MABaI$_3$ shows promising transport properties such as higher carrier concentration, lower effective mass and increasing electrical conductivity with increasing La-doping. MA(Ba$_{1-x}$La$_x$)I$_3$ also shows favorable optical properties, including lower reflectivity and absorption coefficients, favorable for its use as TC applications. Over all, we propose MA(Ba$_{1-x}$La$_x$)I$_3$ to be a good bulk material for TC applications.

## Acknowledgements

AA acknowledges financial support from IIT Bombay via the SEED grant, project code 13IRCCSG020.

## References

1. Xinge Yu, Tobin J. Marks and Antonio Facchetti, Nature Materials 15, 4599 (**2016**).
2. R. G Gordon, MRS Bull. 25, 52 (**2000**).
3. W. J. Yin, J. H. Yang, J. Kang, Y. Yan, and S. H. Wei, J. of Mater. Chem. A 3, 8926 (**2015**).
4. T. J. Jacobsson, M. Pazoki, A. Hagfeldt, and T. Edvinsson, J. Phys. Chem. C 119, 25673 (**2015**).
5. W. Kohn and L. J. Sham, Phys. Rev. 140, A1133 (**1965**).
6. G. Kresse and J. Furthmueller, Comput. Mater. Sci. 6, 15 (**1996**).
7. G. Kresse and D. Joubert, Phys. Rev. B 59, 1758 (**1999**).
8. P. E. Blöchl, Phys. Rev. B 50, 17953 (**1994**).
9. J. P. Perdew, K. Burke, and M. Ernzehof, Phys. Rev. Lett.77, 3865 (**1996**).
10. Georg. K. H. Madsen, David. J. Singh, Computer Physics Communications, 175, 67-71, **2006**
11. Effective Mass Calculator, A. Fonari, C. Sutton, (**2012**).
12. A. Slassi, S. Naji, A. Benyoussef, M. Hamedoun, A. El Kenz, Journal of Alloys and Compounds 605 (**2014**) 118–123.

Figures :

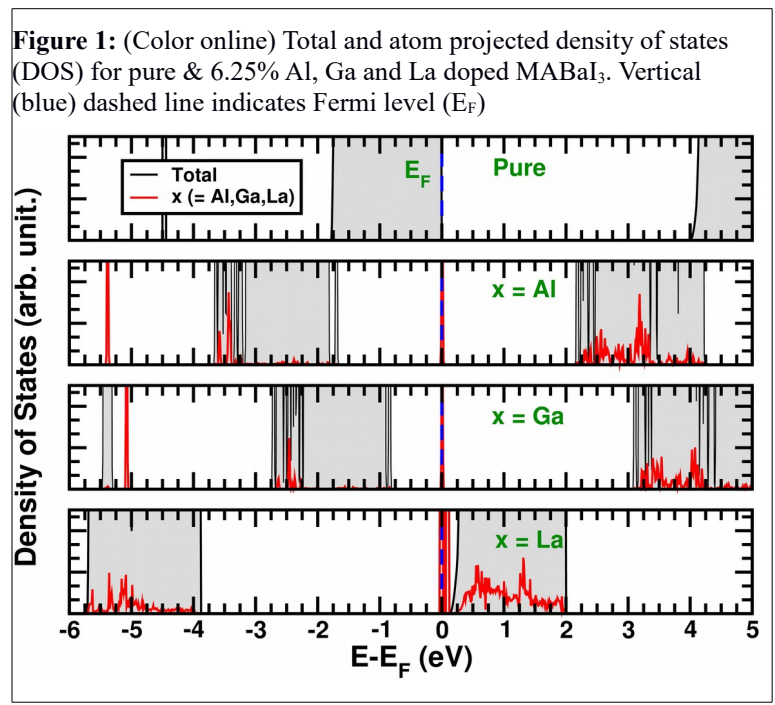

**Figure 1:** (Color online) Total and atom projected density of states (DOS) for pure & 6.25% Al, Ga and La doped MABaI$_3$. Vertical (blue) dashed line indicates Fermi level (E$_F$)

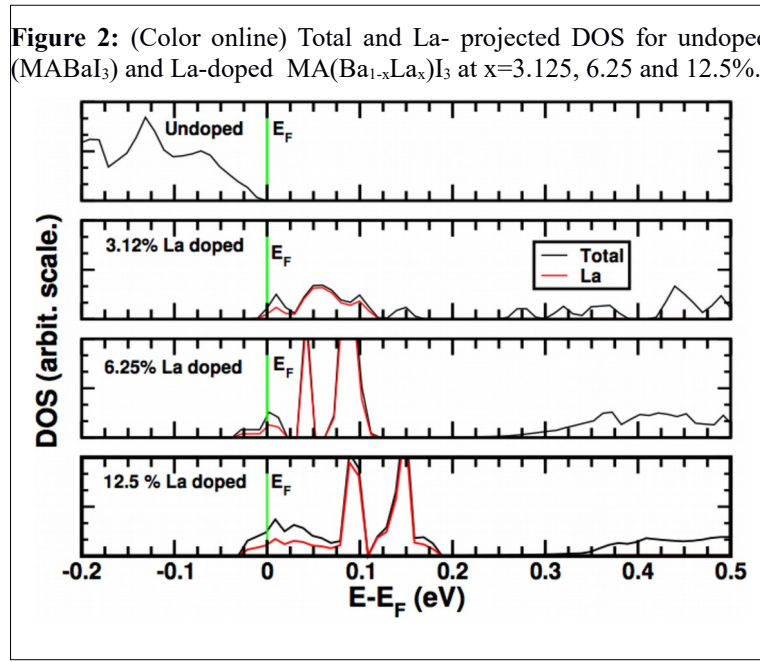

**Figure 2:** (Color online) Total and La- projected DOS for undoped (MABaI$_3$) and La-doped MA(Ba$_{1-x}$La$_x$)I$_3$ at x=3.125, 6.25 and 12.5%.

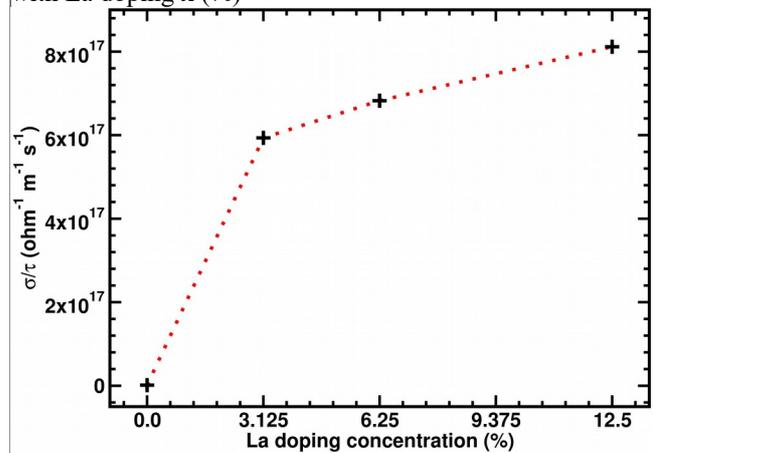
**Figure 3:** (Color online) Variation of electrical conductivity ($\sigma/\tau$) with La doping x (%)

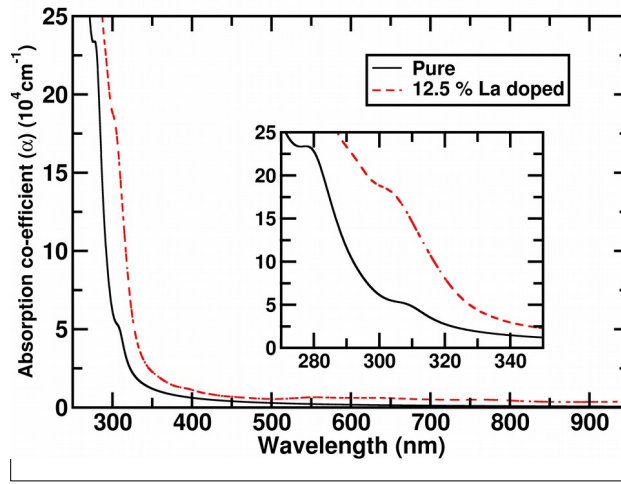

**Figure 4:** (Color online) Optical absorption coefficient for pure and 12.5% La doped MABaI$_3$ in visible region. Inset shows a closer look of peak positions.